# Dissipative phase transition in systems with two-photon drive and nonlinear dissipation near the critical point

V. Yu. Mylnikov*, S. O. Potashin, G. S. Sokolovskii, and N. S. Averkiev

*Ioffe Institute, St. Petersburg, 194021, Russia*
*Corresponding author: vm@mail.ioffe.ru*

(Dated: May 19, 2021)

We study dissipative phase transition near the critical point for a system with two-photon driving and nonlinear dissipation. The proposed mean-field theory, which explicitly takes into account quantum fluctuations, allowed us to describe properly the evolution dynamics of the system and to demonstrate the new effects in the steady-state. We show that the presence of quantum fluctuations leads to a power-law dependence of the anomalous average at the phase transition point, with which the critical exponent is associated. Also, we investigate the effect of the quantum fluctuations on the critical point renormalization and demonstrate the existence of a two-photon pump "threshold". It is noteworthy that the obtained results are in a good agreement with the numerical simulations.

Nowadays driven-dissipative phase transition (DPT) in open quantum systems is among the most rapidly developing fields of quantum optics [1]. DPT can be observed when a direct manipulation of the interaction constants, external driving, or dissipation rates of the system leads to an abrupt and nonanalytical change of the system observables [2,3]. Recent publications report on observation of dissipative critical phenomena and nonequilibrium quantum states in superconducting circuits [4,5], cavity quantum electrodynamics systems [6–8], optomechanical resonators [9], semiconductor microcavities [10], and atomic systems [11–14]. The experimental implementation of highly controllable open nonequilibrium photonic systems became possible due to nonlinear reservoir engineering. It enables realization of the optical cavities with an engineered two-photon drive and nonlinear dissipation [15–20]. In such systems the presence of dissipation does not destroy but stabilizes the quantum state, also known as the Schrödinger cat state [21–23]. Basing on this state, it is possible to prepare a dynamically protected qubit [24] for further applications in quantum information processing [25–30].

The above-mentioned quantum systems are commonly studied with numerical methods. The most notable of these are integration of a master equation on a truncated Fock basis [19] and diagonalization of the Liouvillian superoperator [2,31,32]. Besides, the complex-P-representation [33], Monte Carlo [34], quantum trajectory [35], and the quantum-absorber methods [36] are intensively utilized to investigate the properties of the nonequilibrium stationary state in such problems. However, numerics typically cannot allow to uncover physical phenomena underlying the evolution of the system. This is one of the reasons why the DPT in open quantum systems is not well understood to date. In the limit of a large number of photons, a qualitative picture of the occurring phenomena can be obtained using the so-called semiclassical approximation [37,38]. Nevertheless, this approach completely neglects the quantum fluctuations (QF), which can play a significant role, even in the case of a large average number of photons in the system [10].

In this Letter, we use the formalism of Keldysh Green's functions [39,40], since it explicitly takes into account the effects of QF. Recently it has been shown that the nonequilibrium Keldysh technique is a promising way to study nonequilibrium open quantum systems [41]. It also provides a theoretical framework for the systematic treatment of DPT [41–45]. The application of this approach enables us to construct the self-consistent equations of motion similar to the Gorkov equations in the mean-field approximation. We use the resulting equations to calculate the dynamics of the system observables and to demonstrate the new effects appearing in the steady-state. In analogy with Landau theory, we show that the presence of QF leads to a broadening of the DPT near the critical point. This effect is caused by an "external" QF field. Its existence leads to the power-law dependence of the anomalous average vs the pump rate at the phase transition point with which the critical exponent is associated. In addition, the QF effect causes the critical point renormalization. Both results are in a good agreement with our numerical simulations.

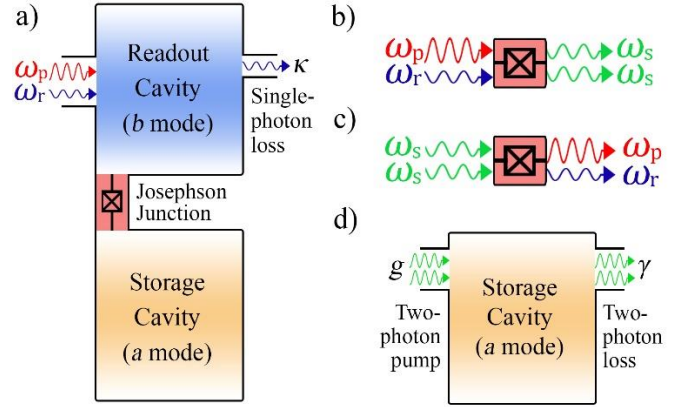

Fig. 1. (a) A schematic of the system. Two superconducting microwave cavities are coupled by a Josephson junction. Pump and drive tones are applied to the readout cavity, which creates a steady-state in the storage cavity. (b-c) Four-wave mixing processes are provided by the presence of nonlinear interaction between the fundamental modes of the readout and storage cavities. One can observe (b) the conversion of the pump and readout photon onto the two storage photons and (c) corresponding backward process. (d) Effective description of the storage cavity: two-photon driving at $g$ rate and two-photon dissipation at $\gamma$ rate (see text for details).

First, we briefly revisit the system described in [24] and experimentally realized in [46]. Its schematic is depicted in Fig. 1 (a), which shows two superconducting microwave cavities. The first resonator has a low Q-factor associated with the presence of significant single-photon losses at $\kappa$ rate. The second resonator has a high Q-factor, so it can be used to store and protect quantum information [23,27]. A transmission line with embedded Josephson junction provides a nonlinear interaction between two cavities, which we denote as readout and storage. In addition, pump and drive microwave tones are applied to the readout cavity at frequencies $\omega_p$ and $\omega_r$. The Hamiltonian of the system has the following form ($\hbar = 1$):

$$H = H_0 + H_{drive} + H_{int}, \quad (1)$$

where $H_0$ – is the Hamiltonian of two linear cavities, $H_{drive}$ represents resonant coherent drive of the readout cavity, as shown in Fig. 1 (a) and $H_{int}$ describes generation of the two identical storage photons from the readout and pump photons, as shown in Fig. 1 (b), and the corresponding backward process, which is shown in Fig. 1 (c). It is important to note that nonresonant coherent drive is embedded into $H_{int}$ [46]. Thus, $H_0 = \omega_s a^\dagger a + \omega_r b^\dagger b$, where $\omega_s$ and $\omega_r$ are the storage and readout frequencies respectively, $a/a^\dagger$ and $b/b^\dagger$ are the annihilation/creation operators corresponding to the fundamental cavity modes. The Hamiltonian of a resonant coherent drive with amplitude $\varepsilon_r$ and frequency $\omega_r$ has the form: $H_{drive} = \varepsilon_r b^\dagger \exp(-i\omega_r t) + \varepsilon_r^* b \exp(i\omega_r t)$. The last term in the equation (1) is determined by the following expression:

$$H_{int} = -\tfrac{1}{2}\mu(\xi e^{-i\omega_p t} b a^{\dagger 2} + \xi^* e^{i\omega_p t} b^\dagger a^2), \quad (2)$$

where $\mu$ is a nonlinear coupling constant emerging from the presence of a Josephson junction [47]. In equation (2) pump photons are considered in the classical field approximation with effective amplitude $\xi = -i\varepsilon_p/[\kappa/2 + i(\omega_r - \omega_p)]$, where $\varepsilon_p$ and $\omega_p$ are amplitude and frequency of the external nonresonant coherent pump [46].

Since dissipative processes play a crucial role in the considered system, its behavior must be described by Lindblad master equation [48]:

$$\partial_t \rho = -i[H, \rho] + \kappa D[b](\rho), \quad (3)$$

where $\rho$ is a system density matrix, $\kappa$ is a single-photon loss rate, $D[b]$ is a Liouvillian, which defines as $D[b](\rho) = b\rho b^\dagger - \tfrac{1}{2}(b^\dagger b \rho + \rho b^\dagger b)$. Notably, Liouvillian acts on the density matrix $\rho$ from both sides and the operator $b$ is called Lindblad operator or a quantum jump operator.

Further, we assume that the characteristic time scale of the single-photon dissipation ($1/\kappa$) is much smaller than all other time scales of the system. Consequently, we can eliminate degrees of freedom associated with the readout cavity [46]. Furthermore, it is convenient to use a unitary transformation $U = \exp[-i(\omega_p + \omega_r)a^\dagger a/2]$, which makes the effective Hamiltonian time-independent. As a result, the following description of the reduced density matrix $\rho_s = \text{Tr}[\rho]_r$ can be obtained:

$$\partial_t \rho_s = -i[H_{eff}, \rho_s] + 2\gamma D[a^2](\rho_s), \quad (4)$$

where the effective Hamiltonian is given by:

$$H_{eff} = \Delta a^\dagger a + i\tfrac{1}{2}(g\, a^{\dagger 2} - g^* a^2), \quad (5)$$

where $g = 2\xi \mu \varepsilon_r / \kappa$ is a two-photon pump rate, $\gamma = |\xi \mu|^2 / 2\kappa$ is a two-photon dissipation rate, and $\Delta = (2\omega_s - \omega_p - \omega_r)/2$ is a frequency detuning. The most interesting is the regime where $\omega_p \approx 2\omega_s - \omega_r$ and hence $\Delta \ll \omega_p, \omega_s, \omega_r$. Further, we suppose that the two-photon pump rate $g$ is a real value. This can be easily achieved by tuning the complex phase of the nonresonant coherent pump amplitude.

It is worth mentioning that $g$ is proportional to the product of the nonresonant pump and resonant drive amplitudes $\xi$ and $\varepsilon_r$. However, the absorption rate is $\gamma \propto \xi^2$ and does not depend on the drive amplitude $\varepsilon_r$. As a result, a two-photon pump and nonlinear dissipation are effectively implemented in the storage cavity due to the presence of a linear dissipation and coherent pump in the readout cavity, as shown in Fig. 1 (d).

From now on, we will consider the behavior of the following two system observables:

$$G^K(t,t) = 2n(t) + 1, \quad F^K(t,t) = 2\psi(t), \quad (6)$$

where $n(t) = \text{Tr}[a^\dagger a \rho_s(t)]$ is an average number of photons in the storage cavity, $\psi(t) = \text{Tr}[a^2 \rho_s(t)]$ is an anomalous average of the system, $G^K(t,t)$ and $F^K(t,t)$ are the normal and anomalous simultaneous Keldysh Green's functions. The unity term in the normal Keldysh Green's function takes into account the presence of the QF [40,41]. As a result, one can obtain equations of motion for simultaneous Keldysh Green's functions from the equation (4):

$$\partial_t G^K(t,t) = gF^K(t,t)^* + g^* F^K(t,t) - 8\gamma \langle a^{\dagger 2} a^2 \rangle,$$
$$(\partial_t + i2\Delta)F^K(t,t) = 2gG^K(t,t) - 4\gamma \langle a^2(2a^\dagger a + 1) \rangle, \quad (7)$$

where angle brackets denote the quantum mechanical averaging over the density matrix of the system. As can be seen from the equation (7), the normal and anomalous Keldysh Green's functions are expressed through the expectation values of higher-order operators, due to the nonlinear dissipation arising in the system. We use the mean-field approximation to avoid considering higher-order equations of motion. Within this approximation, the expectation values of the operator products are replaced by the products of their expectation values [49]. However, there are several ways to implement this decoupling [50]. The two most common choices are as follows:

$$\langle a^{\dagger 2} a^2 \rangle \approx \psi^* \psi, \quad (8a)$$

$$\langle a^{\dagger 2} a^2 \rangle \approx \psi^* \psi + 2n^2, \quad (8b)$$

where in (8a) decoupling was carried out using only the so-called

"Cooper" or "pairing" channel [51]. In (8b) an additional "density" channel appears [51]. Here, we assume (8a) to be valid for the considered system. Justification of this choice will be given in the following text. As a result of (8a), the expectation values of the higher-order operators in the equations (7) can be expressed through Keldysh Green's functions as follows:

$$\langle a^{\dagger 2} a^2 \rangle \approx \tfrac{1}{4} F^K(t,t)^* F^K(t,t),$$
$$\langle a^2 (2a^\dagger a + 1) \rangle \approx \tfrac{1}{2} F^K(t,t) G^K(t,t), \quad (9)$$

and the mean-field self-consistent equations of motion can be obtained:

$$\partial_t G^K(t,t) = g_{\text{eff}}(t) F^K(t,t)^* + g_{\text{eff}}(t)^* F^K(t,t),$$
$$(\partial_t + i2\Delta) F^K(t,t) = 2 g_{\text{eff}}(t) G^K(t,t), \quad (10)$$

where $g_{\text{eff}}(t) = g - 2\gamma \psi(t)$ is a renormalized two-photon pump rate. We will also assume that the initial state is a vacuum, which gives the following initial conditions: $G^K(0,0)=1$ и $F^K(0,0)=0$. The unity term in $G^K$ emerges from the QF, as mentioned above. This enables the parametric generation of light even if there are no photons in the initial state [52]. It should also be noted that the resulting equations (10) have a structure similar to the Gorkov equations, which play a central role in the theory of superconductivity. Another point worth mentioning is that the (10) has the following integral of motion: $G^K(t,t)^2 - |F^K(t,t)|^2$. Thus, it is possible to express the normal Green's function through the anomalous one, using vacuum initial conditions: $G^K(t,t)=(1+|F^K(t,t)|^2)^{1/2}$. In result, one can obtain the equation of motion solely for the anomalous average:

$$\partial_t \psi(t) = -i2\Delta \psi(t) + \sqrt{1 + 4|\psi(t)|^2}\,[g - 2\gamma \psi(t)], \quad (11)$$

and relate to the average number of photons:

$$n(t) = \tfrac{1}{2}\left(\sqrt{1 + 4|\psi(t)|^2} - 1\right). \quad (12)$$

We use the resulting equations of motion of the proposed mean-field theory to calculate the time evolution of the anomalous average $\psi(t)$ and the average number of photons $n(t)$. A comparison of the time evolution obtained from the equations (11)–(12) and the numerical simulation of the master equation [53,54] is shown in Fig. 2 (a-b). It is clear that the mean-field theory qualitatively well describes the evolution of calculated system observables. As pointed above, the influence of QF is tremendous in a near-zero time region by analogy with the parametric generation of light from a vacuum state. As can be seen from Fig. 2, the semiclassical solution is exact zero, since generation cannot start without QF. At longer time scale, the presence of QF significantly affects the dynamics of the system observables, which leads to the formation of a stationary state.

For the case of zero frequency detuning, the following mean-field stationary solutions can be found from Eqs. (11) for the anomalous average and the average number of photons in the storage cavity:

$$\psi = g/2\gamma, \quad n = \tfrac{1}{2}\left(\sqrt{1 + 4\psi^2} - 1\right). \quad (13)$$

In addition, using the exact stationary solution for the density matrix for $\Delta = 0$ [55], one can find $\psi$ and $n$ in the explicit form:

$$\psi = g/2\gamma, \quad n = \psi \tanh(\psi) \quad (14)$$

A comparison of the stationary solutions calculated by the mean-field theory (13) and the exact stationary density matrix (14) demonstrates very good agreement of the anomalous averages, as shown in Fig. 2 (a). However, it can be seen, that the average number of photons differs between the mean-field and exact solutions (Fig. 2 (b)). Nevertheless, both solutions have the same asymptotic behavior in the limit of small ($g \ll \gamma$) and large ($g \gg \gamma$) two-photon pump rate.

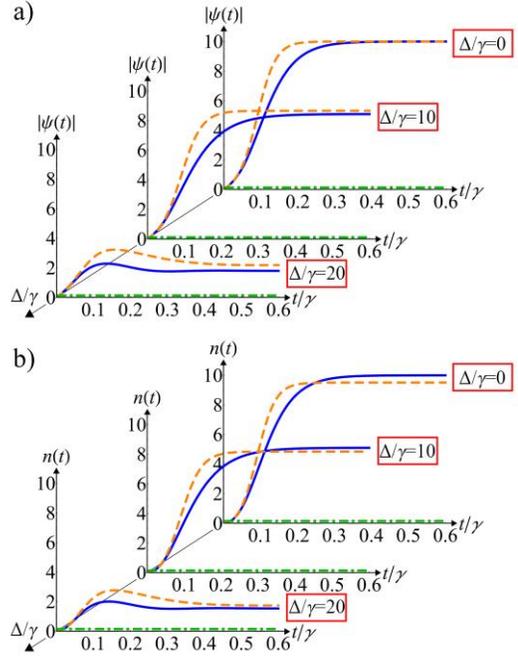

Fig. 2. Time evolution of the modulus of anomalous average $|\psi(t)|$ (a) and the average number of photons $n(t)$ (b) for different values of the frequency detuning $\Delta$. It is obtained from numerical simulation of the Lindblad master equation on a truncated Fock basis (blue curve), numerical integration of the mean-field equations (11)–(12) (orange dashed curve) and the semiclassical solution (green dash-dotted line). The normalized two-photon dissipation rate is set to $g/\gamma = 20$.

For the case of non-zero frequency detuning, we derive the following biquadratic equation on the modulus of the anomalous average:

$$\Delta^2 |\psi|^2 = \tfrac{1}{4}(1 + 4|\psi|^2)(g^2 - 4\gamma^2 |\psi|^2), \quad (15)$$

where the unity term corresponds to the presence of QF. Dropping QF in equation (15), one gets a semiclassical solution [33]:

$$|\psi| \approx \theta(g^2 - \Delta^2)\sqrt{g^2 - \Delta^2}/2\gamma, \quad (16)$$

where $\theta(x)$ is a Heaviside step function. Equation (16) yields non-zero anomalous average $\psi$ when the frequency detuning is below than the two-photon pump rate ($\Delta < g$), and zero $\psi$ value after passing the critical point $\Delta = g$ (green curve in Fig. 3 (b)). For open quantum systems, this phenomenon is also known as DPT [3,33,56,57], and was studied for a considered system earlier [2,33]. Here, we can observe DPT in the "thermodynamic limit", when the average number of photons $n$ or the anomalous average $\psi$ tends to infinity, and one can usually ignore QF [57]. For our system, this limit can be realized in the case of a large two-photon pump rate ($g \gg \gamma$).

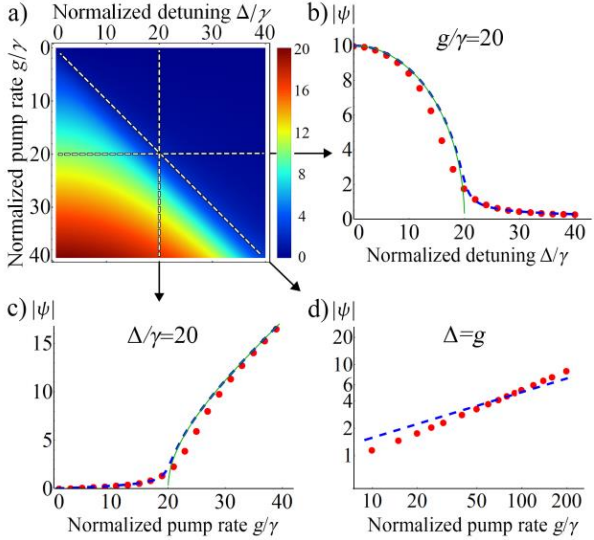

Fig. 3. (a) Phase diagram of the modulus of anomalous average $|\psi|$ as a function of the frequency detuning $\Delta$ and the pump rate $g$ in units of the two-photon dissipation rate $\gamma$. The color plot is computed from the numerical solution of the Lindblad master equation on a truncated Fock basis. (b-c) Corresponding cross-sections at a fixed value of the pump rate $g/\gamma = 20$ (b) and frequency detuning $\Delta/\gamma = 20$ (c). (d) Double logarithmic plot of the modulus of anomalous average $|\psi|$ vs pump rate $g$ at the critical point $\Delta = g$ for $g \gg \gamma$. The calculation was carried out by numerical simulation of the master equation (red dots) and from the mean-field analytic solution (17) (blue dashed line), as well as from the semiclassical approximation (16) (green curve).

We will consider the situation when the average number of photons $n$ and the anomalous average $\psi$ are the finite quantities and it is absolutely necessary to take into account the effects of QF. Within the framework of Keldysh formalism, this leads to the following modification of the semiclassical solution (16):

$$|\psi| = \left( \sqrt{g^2\gamma^2 + \tfrac{1}{4}(g^2 - \Delta^2 - \gamma^2)^2} + \tfrac{1}{2}(g^2 - \Delta^2 - \gamma^2) \right)^{1/2} / 2\gamma. \quad \textbf{(17)}$$

As can be seen from Fig. 3 (a-c), the presence of QF makes the DPT wider near the critical point. This fact is well described by the proposed mean-field theory. To obtain a simple physical explanation for this broadening, we shall demonstrate an analogy with Landau theory of phase transitions [58]. For that, it is necessary to rewrite the equation (15), sorting the contributions in descending order of powers of $\psi$:

$$4\gamma^2 |\psi|^4 + (\Delta^2 - g^2 + \gamma^2)|\psi|^2 = \tfrac{1}{4}g^2, \quad \textbf{(18a)}$$

$$4B \cdot \eta^3 + 2A \cdot \eta = h, \quad \textbf{(18b)}$$

where $\eta$ is an order parameter, $h$ is an external field, $A$ and $B$ are series expansion coefficients of the thermodynamic potential $\Phi = A\eta^2 + B\eta^4 - h\eta$. Minimization of the $\Phi$ yields the expression in (18b). Comparing (18a) and (18b), we can conclude that the presence of QF leads to the analog of the external field $h$. The presence of such an external field in Landau theory breaks the symmetry of the system. As a result, the difference between the two phases disappears, as well as the discrete phase transition point [58]. Therefore, phase transition itself is broadening, which is observed in our system.

Also, it is well known that at the phase transition point defined by a condition $A=0$, the response to the external field is nonlinear and determined by the power-law $\eta \propto (h/B)^{1/\delta}$ [58]. Here $\delta$ is one of the critical exponents for the nonzero external field and it is equal to $\delta = 3$ in Landau theory. Thus, a similar phenomenon should be observed in our system as well. To confirm this statement, we examine the behavior of the anomalous average $\psi$ as a function of the pump rate $g$ at the critical point $\Delta = g$. Here we are interested in the regime $g \gg \gamma$. Fig. 3 (d) demonstrates a comparison between the predictions of the mean-field theory and the exact numerical simulation, which is governed by the following power-law $|\psi| \propto (g/\gamma)^{1/\delta}$ and the corresponding critical exponent: $\delta = 1.47$ for the numerical solution and $\delta = 2$ for the mean-field theory.

It is important to note that the experimental verification of the predicted power-law is very feasible within the framework of the discussed setup (Fig. 1). The frequency detuning $\Delta$ can be controlled by changing the nonresonant pump frequency $\omega_p$, and the two-photon pump rate $g$ by increasing the amplitude $\varepsilon_p$ of the external nonresonant coherent pump wave.

Second non-trivial result associated with QF in the proposed mean-field theory is the renormalization of the critical point. Setting the coefficient before $|\psi|^2$ in (18a) to zero yields the following expression for the case of $g \gg \gamma$:

$$\Delta_0(g) = \sqrt{g^2 - \gamma^2}. \quad \textbf{(19)}$$

The dependence of the critical point on a pump rate $g$ obtained from the mean-field theory (19) is shown in Fig. 4 (a). As one can see, the mean-field theory predicts the existence of a threshold pump rate, which is equal to $g_{\text{th}} = \gamma$.

To understand the physical meaning of the critical point renormalization, it is necessary to consider the case of a small pump rate ($g \ll \gamma$). In this limit, the anomalous average $\psi$ is quite small and one can neglect the quartic part in the equation (18a). Thus, the response of the system to the fluctuation field is determined only by a quadratic contribution. The crucial fact here is that the coefficient before $|\psi|^2$ is a positive quantity for arbitrary frequency detuning $\Delta$ and $g < \gamma$. In Landau theory, such behavior corresponds to the symmetric phase of the system ($A>0$). Further increasing the pump rate $g$ would decrease the corresponding coefficient $(\Delta^2 - g^2 + \gamma^2)$. As result, it becomes zero at the threshold point $g_{th} = \gamma$ and $\Delta = 0$, which violates the linear response and makes the quartic coefficient dominant. This transformation of the system's response correlates precisely with the emergence of the discussed critical behavior.

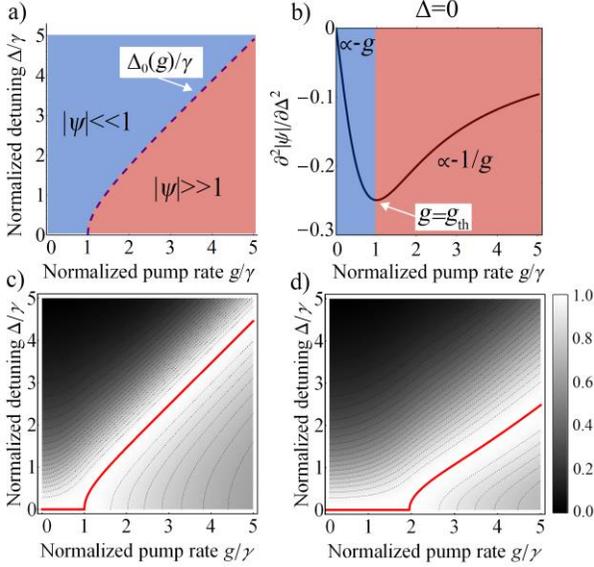

Fig. 4. (a) Phase diagram of the modulus of anomalous average $|\psi|$ in 2D space of the two-photon pump rate $g$ and the frequency detuning $\Delta$ in units of the two-photon dissipation rate $\gamma$, obtained from the mean-field theory. Critical boundary $\Delta_0(g)$ (blue dashed line) separates the phases with low (blue area) and large (red area) average number of photons. (b) The behavior of the second derivative $\partial^2|\psi|/\partial\Delta^2$ vs the pump rate $g$ for $\Delta = 0$. One can see the existence of two regimes, in which switching occurs after passing the threshold point $g_{th} = \gamma$. (c-d) Heat maps of the normalized susceptibility $\chi = \partial|\psi|/\partial g$ as a function of a normalized pump rate $g$ and frequency detuning $\Delta$, obtained from (c) mean-field analytic solution (17) and (d) numerical simulation of the master equation. The red line identifies the location of the maximum susceptibility and indicates that the critical behavior occurs only after passing the cutoff pump rate $g_{th}$. The susceptibility $\chi$ is normalized by the maximum value for each frequency detuning $\Delta$.

One can observe such a transition from the analysis of the second derivative $\partial^2|\psi|/\partial\Delta^2$ in the region $\Delta \approx 0$, which is shown in Fig. 4 (b). It can be seen that its behavior changes from the linear asymptotic $\propto g$ to the inverse proportion $\propto 1/g$ after passing the threshold point $g_{th} = \gamma$. Furthermore, considering the dynamics of the system's susceptibility $\chi = \partial|\psi|/\partial g$ as a function of the pump rate $g$ and frequency detuning $\Delta$, we also observe the existence of a threshold pump $g_{th}$. Its behavior in the framework of the mean-field theory is shown in Fig. 4 (c). One can see that for $g < \gamma$ the maximum susceptibility is at the point $\Delta_{max} = 0$.

However, after passing the threshold point $g_{th} = \gamma$ one can observe a shift of the maximum, which indicates the emergence of the critical behavior. The numerical simulations of the system's susceptibility shown in Fig. 4 (d) also predict the existence of $g_{th} \approx 1.952\gamma$, which is in qualitative agreement with the proposed Keldysh formalism.

Finally, justification and constraints of our mean-field treatment should be discussed. It is well known that the DFT in the considered system is associated with transformation from the Schrödinger cat to the squeezed-vacuum state [33] as shown in Fig. 3 (b). For both states, the anomalous average $\psi \neq 0$, but they differ in the decoupling of channels, as shown in expressions (8a) and (8b). As it turns out, the proposed decoupling is rigorous only for a large number of photons region, which corresponds to the cat state. However, in the region of a small number of photons, corresponding to the squeezed state, the anomalous average $|\psi| << 1$. From equation (12) follows that: $|\psi| \gg n$. Consequently, in the region of a small number of photons, the "Cooper" channel dominates over the "density" channel and the proposed decoupling is valid. Also, we have shown above that near the critical point, the proposed description allows us to achieve a good qualitative agreement with the exact results of numerical calculations. We believe that the observed mismatch can be explained by the presence of a fluctuation region, where the QF significantly affects the behavior of the system. Thus, for further exploration, it seems to be necessary to apply more precise methods, such as the Keldysh functional integral [41], the 2PI effective action [59], and the renormalization group [60,61].

*Conclusions and outlook.*—In this Letter, we develop a mean-field theory for a system with two-photon driving and dissipation, which explicitly takes into account the quantum fluctuations. Consideration of quantum fluctuations allows us to describe properly the evolution dynamics of the system and to demonstrate the new effects in the steady-state. We show that the dissipative phase transition broadening near the critical point is naturally conditioned by the fluctuation field. Its presence leads to a power-law dependence of the anomalous average at the phase transition point, with which the critical exponent is associated. This counterintuitive effect cannot be found in the semiclassical approximation. The reason lies in the crucial role of the quantum fluctuations, which significantly affect the behavior of a given system at the critical point and should be taken into account even in the case of a large average number of photons. Also, we investigate the effect of the quantum fluctuations on the critical point renormalization and demonstrate the existence of a two-photon pump "threshold". It is noteworthy that the obtained results are in a good agreement with the numerical simulations. An important point here is that intuitively free energy analysis is out of consideration for dissipative phase transitions. However, obtained counter-intuitive results are largely based on analogies with Landau theory. We believe that further investigations will clarify this close relationship between purely non-equilibrium formalism and traditional equilibrium techniques. The results presented in this Letter can be applied to the development of the new nonequilibrium quantum states with controllable properties [62,63] for applications in quantum information processing [64] and metrology [65].

We acknowledge fruitful discussions with Mikhail Glazov, Valentin Kachorovskii and Dmitry Mikhailov.